\documentclass[%
 reprint,
 amsmath,amssymb,
 aps,
]{revtex4-2}

\usepackage{graphicx}% Include figure files
\usepackage{dcolumn}% Align table columns on decimal point
\usepackage{bm}% bold math
\begin{document}

\preprint{APS/123-QED}

\title{Pressure and temperature dependence of fluorescence anisotropy of Green Fluorescent Protein}% Force line breaks with \\

\author{Harpreet Kaur}
%\altaffiliation[Also at ]{}%Lines break automatically or can be forced with \\
\author{Khanh Nguyen}%
\author{Pradeep Kumar}
\email{pradeepk@uark.edu}
\affiliation{%
Department of Physics, University of Arkansas, Fayetteville, AR 72701 USA
}%

\date{\today}% It is always \today, today,

\begin{abstract}
We have studied the effect of high hydrostatic pressure and temperature on the steady state fluorescence anisotropy of Green Fluorescent Protein (GFP).  We find that the fluorescence anisotropy of  GFP at a constant temperature decreases with increasing pressure. At atmospheric pressure, anisotropy decreases with increasing temperature but exhibits a maximum with temperature for pressure larger than $20$~MPa. The temperature corresponding to the maximum of anisotropy increases with increasing pressure. By taking into account of the rotational correlation time changes of GFP with the pressure-temperature dependent viscosity of the solvent, we argue that viscosity increase with pressure is not a major contributing factor to the decrease in anisotropy with pressure. The decrease of anisotropy with pressure may result from changes in H-bonding environment around the chromophore.
\end{abstract}

\maketitle
%%%Please don't disable any packages in the preamble, as this may cause the template to display incorrectly.%%%

%%%MAIN TEXT%%%%
Green Fluorescent Protein (GFP) is a cylindrical  protein made up of 238 amino acid residues~\cite{ormo1996crystal, tsien1998green}. It consists of 11 $\beta$-barrels connected by short $\alpha$-helices with an $\alpha$-helix inside the cylinder~\cite{ormo1996crystal, tsien1998green}. A chromophore is formed with cyclization of the  Ser65, Try66 and Gly67 amino acid residues~\cite{tsien1998green, reid1997chromophore}. The GFP chromophore absorbs blue light (a primary peak at $\approx 395$~nm and secondary peak $\approx 480$~nm) and emits green light  with a maximum intensity at about $508$nm~\cite{ward1982spectral}. The chromophore is well centered and connected to the alpha helix inside the $\beta$-barrel can~\cite{ormo1996crystal, tsien1998green}. In its native state, the chromophore is protected from solvent and ions outside the $\beta$-barrel. This unique structure and high fluorescence yield in the native state has made GFP a very popular candidate for various {\it in vivo} and {\it in vitro} measurements to probe the environment of living cells~\cite{tsien1998green, chalfie1994green, zimmer2002green}. 

Due to pH sensitivity of GFP fluorescence, GFP and its mutants have been used as pH indicators for cytosolic, nuclear, and mitocondrial regions in Hela cells~\cite{llopis1998measurement}. In recent studies, measurement of fluorescence anisotropy of GFP is explored under different situations for various purposes. For example, Mattheyses et. al have used fluorescence anisotropy of GFP-tagged nucleoporins to reveal the packing of the proteins within the nuclear pore complex~\cite{mattheyses2010fluorescence}. Donner et. al. have used GFP as an intracellular temperature sensor by measuring temperature dependence of fluorescence anisotropy~\cite{donner2012mapping}. Fluorescence anisotropy is advantageous over fluorescence intensity as it is independent of the factors such as non-uniformity in molecular concentration and intensity of the excitation light~\cite{valeur2012molecular,donner2012mapping}. 

On earth, a large majority of bacteria and archaea grow in a wide array of environmental conditions including high pressures, extremes of temperature, pH, and salinity~\cite{kato1998extremely,brock1969thermus,oshima1974description,takami1997microbial,schleper1995life,maheshwari2015halophiles}. Because these conditions are not hospitable for other life forms, these organisms have been named extremophiles. For example,~\emph{Thermus aquaticus}, a thermophilic bacteria, grows optimally at $70^{\circ}$C~\cite{brock1969thermus}. 
Obligatory barophilic (pressure loving) bacteria from Mariana Trench, the deepest known point ($11$~km) where the hydrostatic pressure can be $\approx110$~MPa, have been isolated and studied~\cite{kato1998extremely}. Recent experiments suggest that even a mesophilic bacterium,~\emph{Escherichia coli}, can grow at pressure as high as $50$~MPa in a temperature dependent manner~\cite{kumar2013pressure,nepal2018dynamics}. Analogs of environmental extremes on earth also exist on other planetary bodies of astrobiological interest. Europa, a moon of Jupiter, is a prime candidate due to the presence of liquid water ocean running $100-200$~km deep, with hydrostatic pressure reaching $130-260$~MPa~\cite{naganuma1998dive}. Therefore, the use of GFP as a cellular probe, a systematic study of the effect of these environmental conditions on the properties of GFP is required.

In the present study, we have investigated the effect of high hydrostatic pressure and temperature on the fluorescence anisotropy of GFP. Specifically, we measure the steady state fluorescence anisotropy of GFP  in a wide range of pressure ($0.1-200$~MPa) and temperature ($10-70^{\circ}$C). We have further analyzed the changes observed in the anisotropy due to  pressure and temperature by taking into account the pressure-temperature dependence of  dynamic viscosity of water. In the {\bf Materials and Methods} section, we describe our experimental setup for high pressure and temperature. In section {\bf Results}, we discuss the results on the pressure and temperature dependence of fluorescence anisotropy of GFP and finally, we conclude with the {\bf Discussion} section.  
\begin{figure*}
  \begin{center}
\includegraphics[width=10cm]{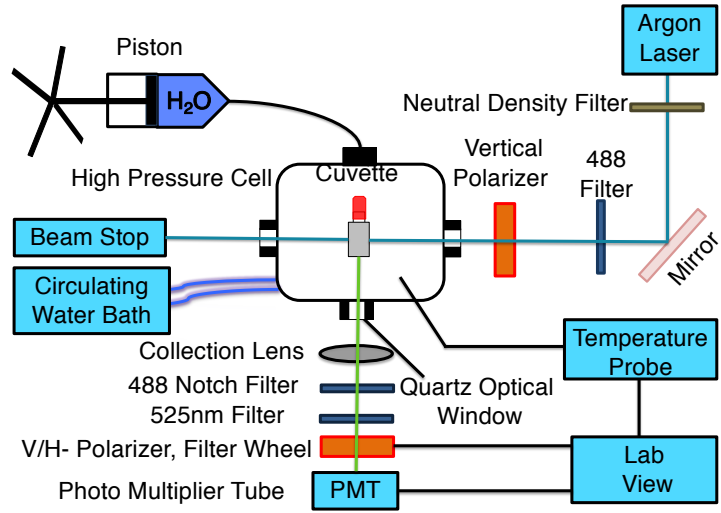}
\end{center}
\caption{Experimental setup for temperature regulated fluorescence anisotropy measurements at different hydrostatic pressures.}
\label{fig:setup}
 \end{figure*}
\section{Materials and Methods}
\subsection{Experimental Setup:} 
 The experimental setup to measure fluorescence anisotropy is shown in Fig.~\ref{fig:setup}. A sample in a square cuvette (Spectrocell; volume: 400~$\mu$L) with a flexible movable rubber cap is loaded into the high pressure cell (ISS, Illinois, USA). A piston (HIP Inc, PA, USA) is used to pressurize the water inside the pressure cell, and the pressure is measured using a pressure gauge~\cite{kumar2013pressure}. Since the compressibility of water is very small, a small change in the volume due to compression of the teflon cap allows to transmit pressure from the surrounding water to the sample inside the cuvette. Temperature of the sample was maintained using a circulating water bath (NesLab, USA), and was measured in real time using a thermocouple (National Instruments, USA) connected to the pressure cell. The temperature fluctuations were of the order of $\pm 0.2^o$C and the pressure uncertainty is estimated to be about $1$~MPa. To excite GFP, light from a continuous wave Argon laser ($488$~nm) (Model: 532-TOPO-A01, Melles Griot, USA) is guided through a linear vertical polarizer using a mirror and is incident on the sample cuvette. The emitted light is then collected at an angle $90^{o}$ to the incident light using a plano-convex lens ($f=60$~mm) (Thorlabs, USA) and a $525\pm25$~nm band pass filter (Chroma, USA). An automated filter wheel (TS103, Thorlabs, USA) containing linear polarizer is then used to select the vertical and horizontal intensities of the emitted light consecutively. Following which, a focusing lens ($f=25$~mm) is used to focus the emitted light onto a photo multiplier tube (Hamamatsu, Japan). The intensity of the vertical and horizontal emitted light is measured using a data acquisition (DAQ) card (National Instrument, USA) equipped with $32$-bit frequency counters. Vertical and horizontal intensity measurements were performed by counting the signals for $2$~sec. For each thermodynamic state point, we obtained $30$ measurements of the steady state fluorescence anisotropy values every $10$~sec. Total measurement time for a state point lasted $5$ minutes. The laser intensity fluctuations were less than $1\%$ between the measurements. The error on the fluorescence anisotropy for each state point is obtained from the standard deviation of these values.
\subsection{Fluorescence Anisotropy:}

When exposed to polarized light, chromophores that have their absorption transition moments oriented along the electric vector of the incident light are preferentially excited. Hence the excited-state population is partially oriented along the electric vector of the polarized excitation light. The rotational diffusion of the fluorophores results in depolarization of the emitted light. The degree of anisotropy of the polarization of the emitted light is described by fluorescence anisotropy, $r$~\cite{lakowicz2013principles}, and is given by 

\begin{equation}
r = \frac{I_{\parallel}-I_{\perp}}{I_{\parallel}+2I_{\perp}}
\end{equation} 
where $I_{\parallel}$ is the emission intensity in the direction of polarization of the excitation light, and $I_{\perp}$ is the emission intensity in a perpendicular direction. The denominator, which is proportional to the total intensity of the emitted light, is used for normalization.

Following Perrin~\cite{perrin1926polarisation}, the fluorescence anisotropy $r$ is given by 
\begin{equation}
r= \frac{r_0}{1+\frac{\tau_F}{\tau_R}}
\label{eq:eqAniso}
\end{equation}
where $\tau_R$ is the rotational correlation time of the molecule and $\tau_F$ is the fluorescence lifetime. The constant $r_0$ is the limiting anisotropy, which is theoretically $0.4$ but depends on the sensitivity of the experimental setup. Using the values $\tau_F=2.5$~ns~\cite{hink2000structural}, $\tau_R=10.6$~ns~\cite{hink2000structural} and experimentally measured value of anisotropy $r=0.28$ at room temperature, we find that  $r_0=0.346$. Fluorescence spectra of GFP were obtained using USB4000 (Ocean Optics, USA) spectrometer.

\subsection{Polarization correction at high pressures:}

It is known the fused quartz can scramble the polarization at high pressure. To correct for this in our experiments, we follow the method as described by Paladini and Weber~\cite{paladini1981,paladini1981A}. We used fluorescein ($1$~nM) in glycerol at room temperature for these experiments. We first calculate the G-factor at atmospheric pressure, defined as
\begin{equation}
G = \frac{I_{HV}}{I_{HH}}
\end{equation}
where $I_{HV}$ is the intensity corresponding to horizontal excitation and vertical emission, and $I_{HH}$ is the intensity corresponding to horizontal excitation and horizontal emission. Following this, for each pressure, we calculate the ratio, $R(p)$, given by
\begin{equation}
R(p) = {\frac{I_{VV}}{I_{VH}}}
\end{equation}
The apparent (measured) polarization at each pressure, $p$, is given by
\begin{equation}
P'(p) = \frac{R(p)/G-1}{R(p)/G-1}
\end{equation}
\begin{figure}
% \begin{center}
\includegraphics[width=8cm]{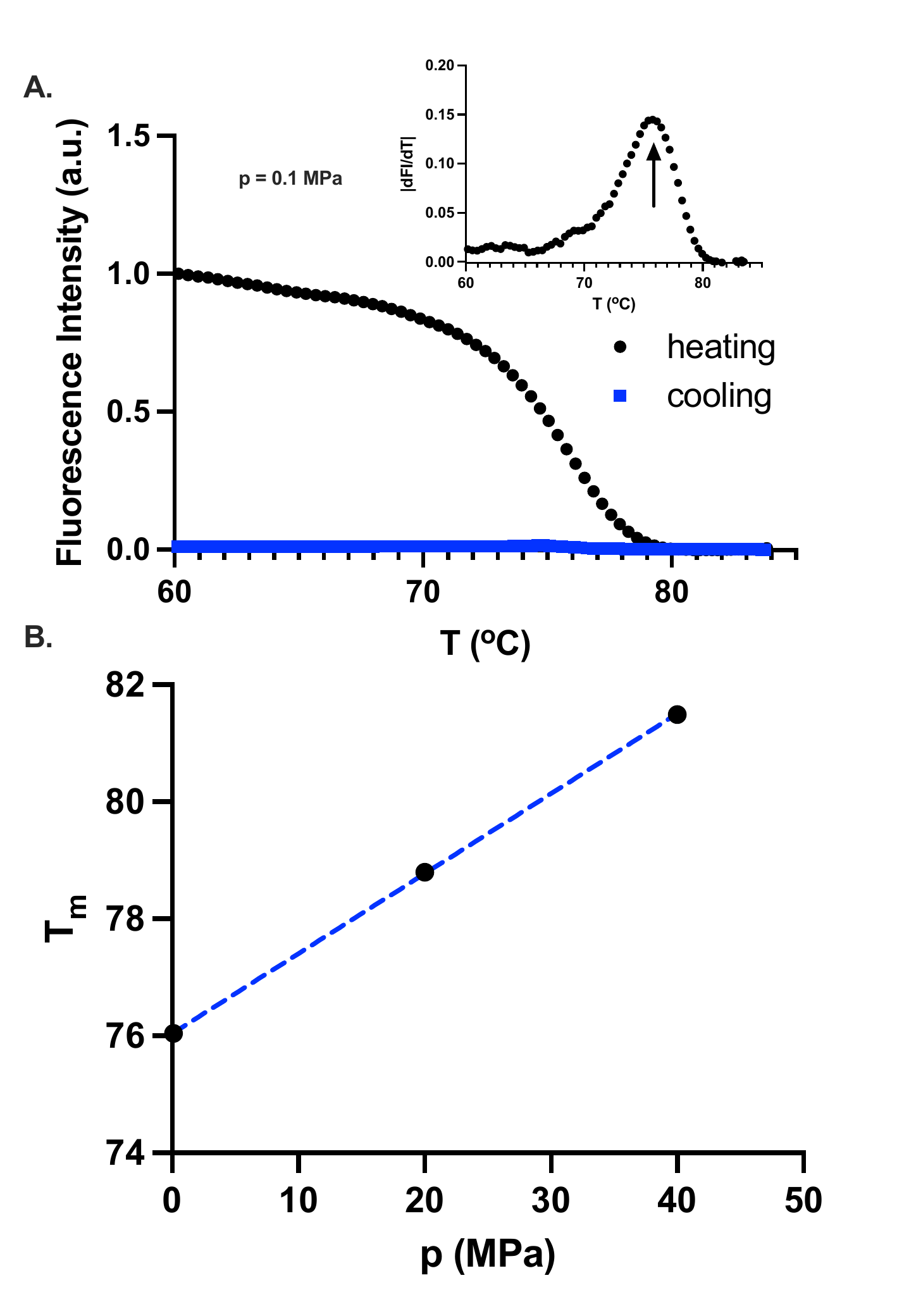}
%\end{center}
\caption{(A) Temperature dependence of Fluorescence intensity at $p=0.1$~MPa during the heating (black circles). Fluorescence intensity during the cooling process is shown as blue squares. Fluorescence of GFP is irreversibly inactivated at high temperatures. Absolute value of the temperature-derivative of fluorescence intensity is shown in the inset. Melting temperature is estimated from the position of maximum of the temperature derivative. (B) Melting temperature, $T_m$, as a function of pressure. Dotted blue curve is a linear fit through the data points. Melting temperature increases with pressure.}
\label{fig:figmelting}
 \end{figure}
Following Ref.~\cite{paladini1981}, the apparent polarization, $P'$, for the L-mode setup (excitation is polarized vertically in respect to the plane of the optics, and anisotropy is calculated from consecutively measured vertical and horizontal polarized intensities) here can be written as   
\begin{equation}
P'(p) = P(1-3\alpha(p))/(1-P\alpha(p))
\end{equation}
where $P$ is the actual polarization and $\alpha$ is a pressure-dependent scrambling coefficient~\cite{paladini1981}. To calculate the correction factor $\alpha$, we use fluorescein at room temperature and calculate $\alpha$ for all the pressure points used in the experiments. Here we have assumed that the polarization scrambling of the excitation and emission lights are the same~\cite{paladini1981}. Assuming that the polarization of the fluorescein in glycerol is pressure-independent~\cite{paladini1981}, $\alpha$ is given by
\begin{equation}
\alpha(p) = \frac{\left [ \frac{P'(p)}{P(0.1\mathrm{MPa})}\right ]-1}{P'(p)-3}
\end{equation}
where $P'$ is the measured polarization at high pressures and $P(0.1\mathrm{MPa})$ is the polarization at $p=0.1$~MPa. In our correction, we further assume that temperature does not affect the $G$ factor, and, therefore, the same value of $G$ was used for all the temperatures.
\subsection {Sample Preparation:}

Wild-type green fluorescent protein (wtGFP) was purchased from Novus Biologicals, USA. All experiments were carried out with $0.5$~$\mu$M of GFP in TMD buffer consisting of 25mM Tris salt (BD Difco, USA), $10$~mM MgCl$_{2}$ (VWR, USA)  and  $1$~mM DTT (VWR, USA). The pH of the sample was adjusted to 7.5$\pm0.1$ using HCl.
\section{Results}
 \begin{figure*}
  \begin{center}
\includegraphics[width=14cm]{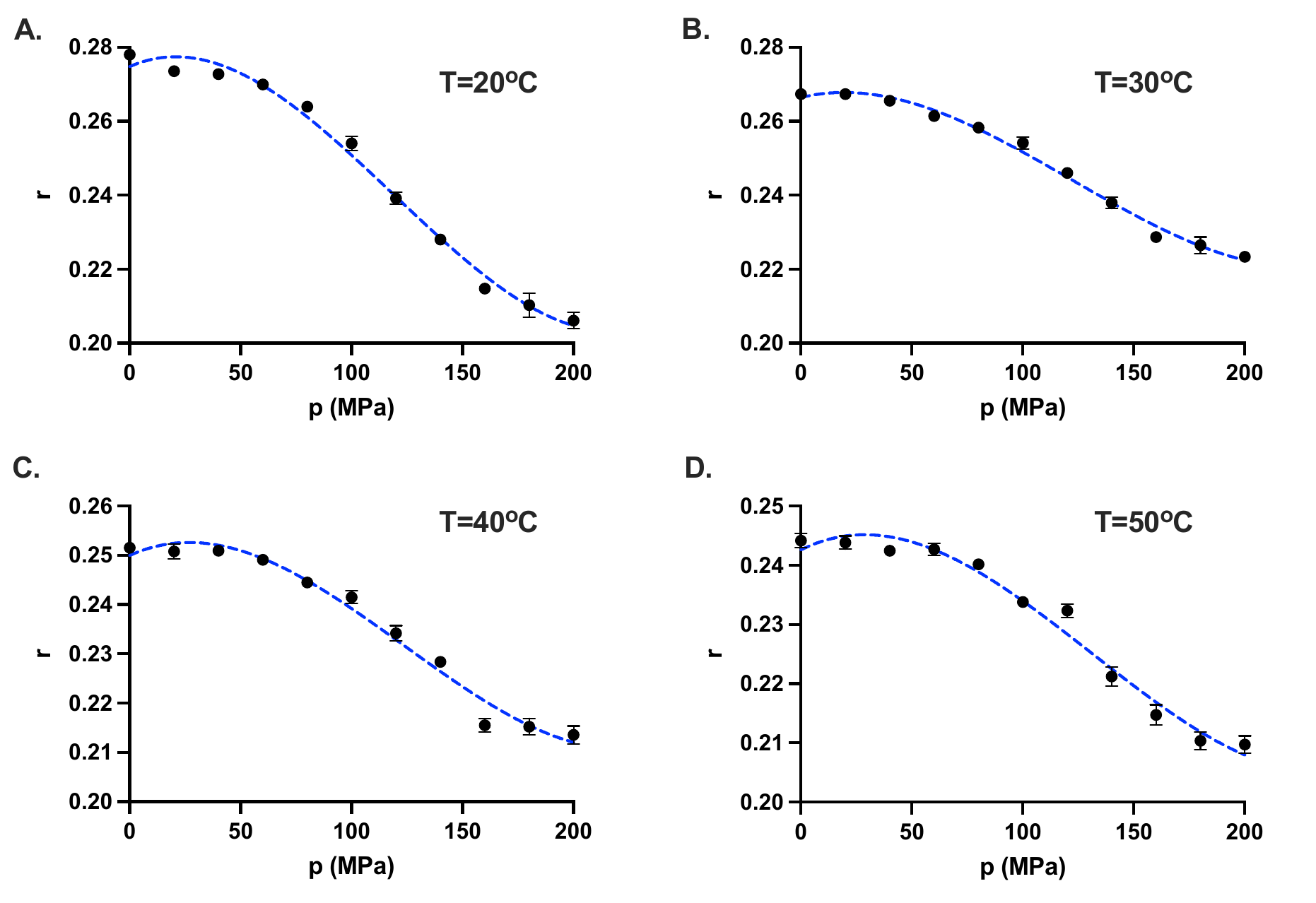}
\end{center}
\caption{Effect of increasing pressure on fluorescence anisotropy of GFP at constant temperature. Pressure dependence of anisotropy is shown for (A) 20$^{\circ}$C , (B) 30$^{\circ}$C , (C) 40$^{\circ}$C , and (D) 50$^{\circ}$C. We also show the error on measurements. Error bars where not visible are smaller than the symbol size. Anisotropy decreases upon increasing pressure for all the temperatures. Dotted blue curves are the polynomial fits through the data points and only serve as the guides to the eye.}
\label{fig:fig3}
\end{figure*}
\subsection{Thermal stability of GFP with pressure:} 
We first determine the thermal stability of GFP for different pressures. For these experiments, $0.5\mu$M GFP in TMD buffer was heated slowly until the fluorescence is inactivated. To determine if the temperature induced fluorescence inactivation is reversible, the sample was cooled following the inactivation. In Fig.~\ref{fig:figmelting}(A), we show the temperature dependence of the fluorescence intensity for atmospheric pressure during the heating (black circles) and cooling process (blue squares). During the heating, fluorescence decreases with temperature and exhibits a sharp transition near the melting temperature. Furthermore, fluorescence does not recover upon cooling, suggesting that the temperature induced denaturation of GFP is irreversible. The melting temperature, $T_m$, is determined from the location of the maximum of the absolute value of the temperature-derivative of fluorescence intensity. The melting temperature was obtained for three pressures, $0.1$~MPa, $20$~MPa, and $40$~MPa. The value of $T_m\approx76^{\circ}$C at atmospheric pressure is similar to the melting temperature reported in Refs.~\cite{ward1982spectral,nicholls2013structural}. Melting temperature increases with pressure as shown in Fig.~\ref{fig:figmelting}(B). Our experimental setup is limited in probing $T_m$ for pressures larger than $40$~MPa as the melting temperature becomes very large ($T_m >82^{\circ}$C). 

We next performed steady state measurements of fluorescence anisotropy at pressures, $p$, between $0.1$~MPa and $200$~MPa at an interval of $20$~MPa, and temperature, $T$, between $10^{\circ}$C and $70^{\circ}$C at an interval of $10^{\circ}$C. Melting temperature of GFP is larger than the range of temperature studied here for all the pressures. Measurements of anisotropy were performed by fixing the temperature and then changing to different values of pressures. Measured anisotropy values had small fluctuations ($\approx \pm 0.002$) around the mean value. An example of experimental data of measured anisotropy values at $p=0.1$~MPa and $T=10^{\circ}$C is shown in Fig.S1 in the Supplementary Materials.

\subsection{Effect of pressure on fluorescence anisotropy at constant temperature:} 
In Fig.~\ref{fig:fig3}, we show the effect of  pressure on the anisotropy of GFP upto $2000$~atm for different constant temperatures $T=10^{\circ}$C, $30^{\circ}$C, $40^{\circ}$C, and $50^{\circ}$C. The error bars are also shown for all the measurements.  For all the temperatures studied here, anisotropy decreases with increasing pressure. Like the fluorescence intensity, fluorescence anisotropy is completely reversible for pressure up to $p=200$~MPa. The change in anisotropy between atmospheric pressure and $200$~MPa is largest for the lowest temperature. 
\begin{figure*}
\begin{center}
\includegraphics[width=16cm]{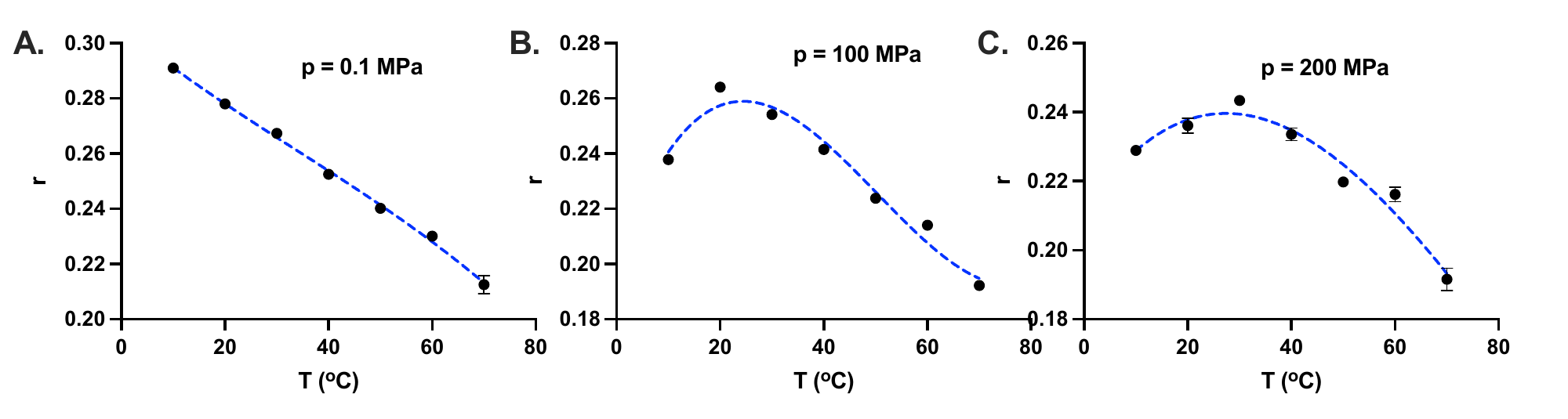}
\end{center}
\caption{Effect of increasing temperature on fluorescence anisotropy of GFP at constant pressure. Temperature dependence of anisotropy is shown for (A) $p=0.1$~MPa, (B) $p=100$~MPa, and (C) $p=200$~MPa.  We also show the error bars for all the measurements. Error bars where not visible are smaller than the symbol size. For atmospheric pressure, anisotropy is a monotonically decreasing function of temperature. Anisotropy exhibits a non-monotonic behavior with temperature with a maximum for $p\geq 20$~MPa. The temperature corresponding to the maximum of anisotropy increases upon increasing pressure. Dotted blue curves are the polynomial fits through the data points and only serve as the guides to the eye.}
\label{fig:fig4}
\end{figure*}
\subsection{Effect of temperature on fluorescence anisotropy at constant pressure:} 
Figure~\ref{fig:fig4} represents the behavior of anisotropy with temperature for fixed pressures (A) $p=0.1$~MPa, (B) $p=100$~MPa, and (C) $p=200$~MPa. For $P=0.1$~MPa, anisotropy decreases with increasing temperature. But for $P\geq20$~MPa, anisotropy exhibits a maximum with temperature. The temperature corresponding to the maximum of anisotropy increases upon increasing pressure. Since the dynamic viscosity of solvent decreases with increasing temperature, one may expect the anisotropy to increase with decreasing temperature. This suggests that the decrease of anisotropy upon lowering temperature at high pressures arises from structural changes of GFP at lower temperatures, reminiscent of cold unfolding of protein~\cite{privalov1990cold,buldyrev2007water}. Even though, the cold denaturation of GFP does not occur at $10^{\circ}$C and pressures as low as $p=20$~MPa, our data suggest that the anisotropy reflects either the structural changes or changes in hydrogen-bonding environment around the chromophore upon further lowering of temperature.

\subsection{Pressure-temperature dependence of anisotropy:} 
\begin{figure}
\begin{center}
 \includegraphics[width=8cm]{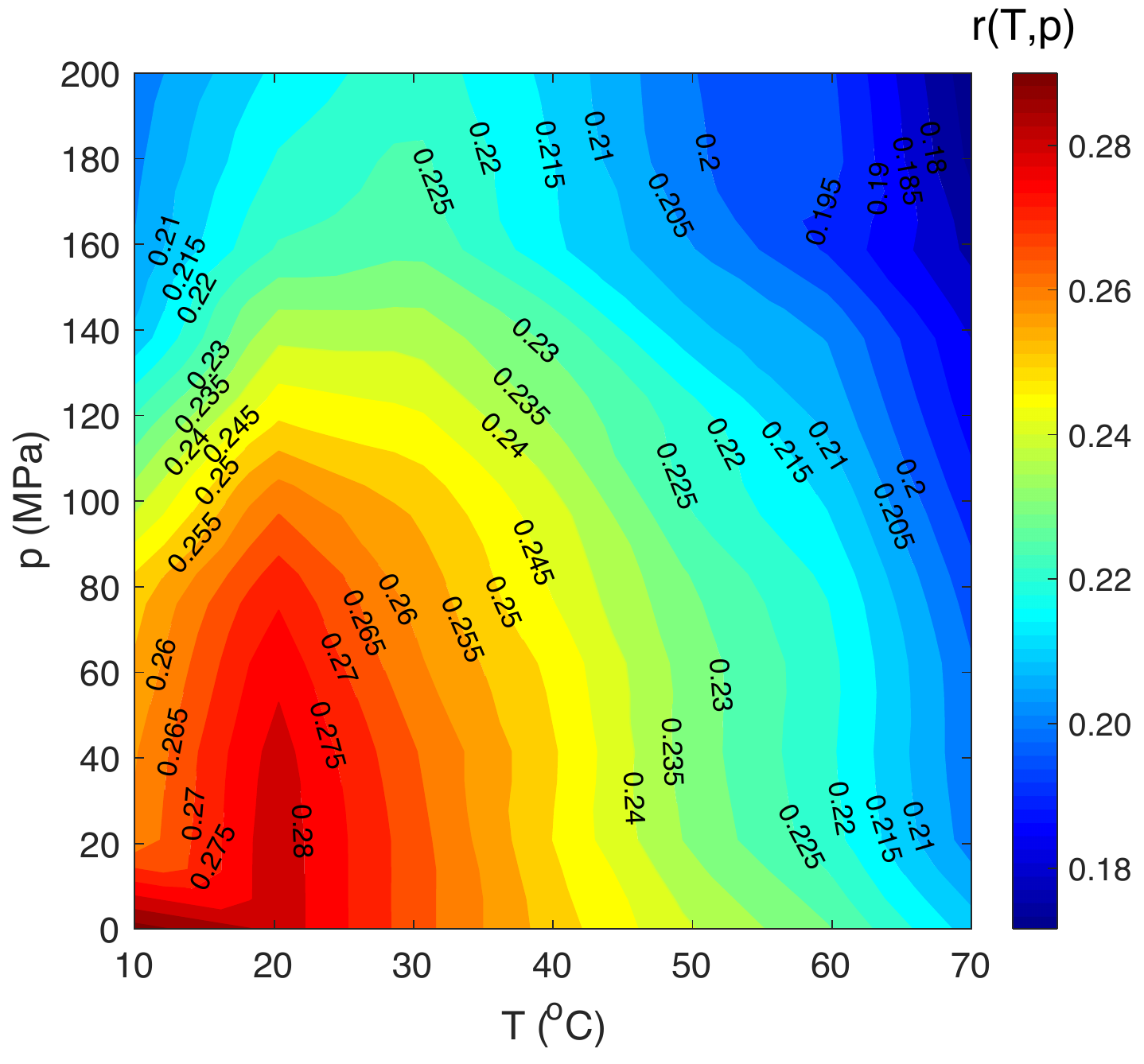}
\end{center}
\caption{P-T surface plot of the fluorescence anisotropy of GFP. Also shown are isoanisotropic contours. Isoanisotropic contours exhibit elliptic shapes typical of P-T phase diagram of protein stability.}
\label{fig:fig5}
 \end{figure}

In Fig.~\ref{fig:fig5}, we show the pressure-temperature surface plot of fluorescence anisotropy of GFP obtained from equilibrium measurements at all the $77$~state points. The isoanisotropic  contours (pressure-temperature curves where anisotropy is a constant) are also shown in Fig.~\ref{fig:fig5}. Isoanisotropic contours exhibit elliptic shapes very typical of pressure-temperature phase diagram of protein stability~\cite{hawley1971reversible}. It is worthwhile to note that the temperature dependence of fluorescence intensity for $p\ge20$~MPa does not exhibit a maximum and decreases monotonically with temperature. The elliptic shape describing the phase boundary in the case of protein denaturation arises due to specific relation between the changes in isothermal compressibility, constant pressure specific heat, and coefficient of thermal expansion between the native and the denatured state~\cite{hawley1971reversible,lesch2004protein,wiedersich2008temperature}. Our experimental data suggest that GFP does not denature in the pressure and temperature range studied here. One possible reason for the elliptic shape of the isoanisotropic contours could be the pressure-temperature dependent changes in hydrogen-bond environment around the chromophore. 
\subsection{\bf Effect of viscosity on anisotropy:} 
\begin{figure}
\begin{center}
\includegraphics[width=8cm]{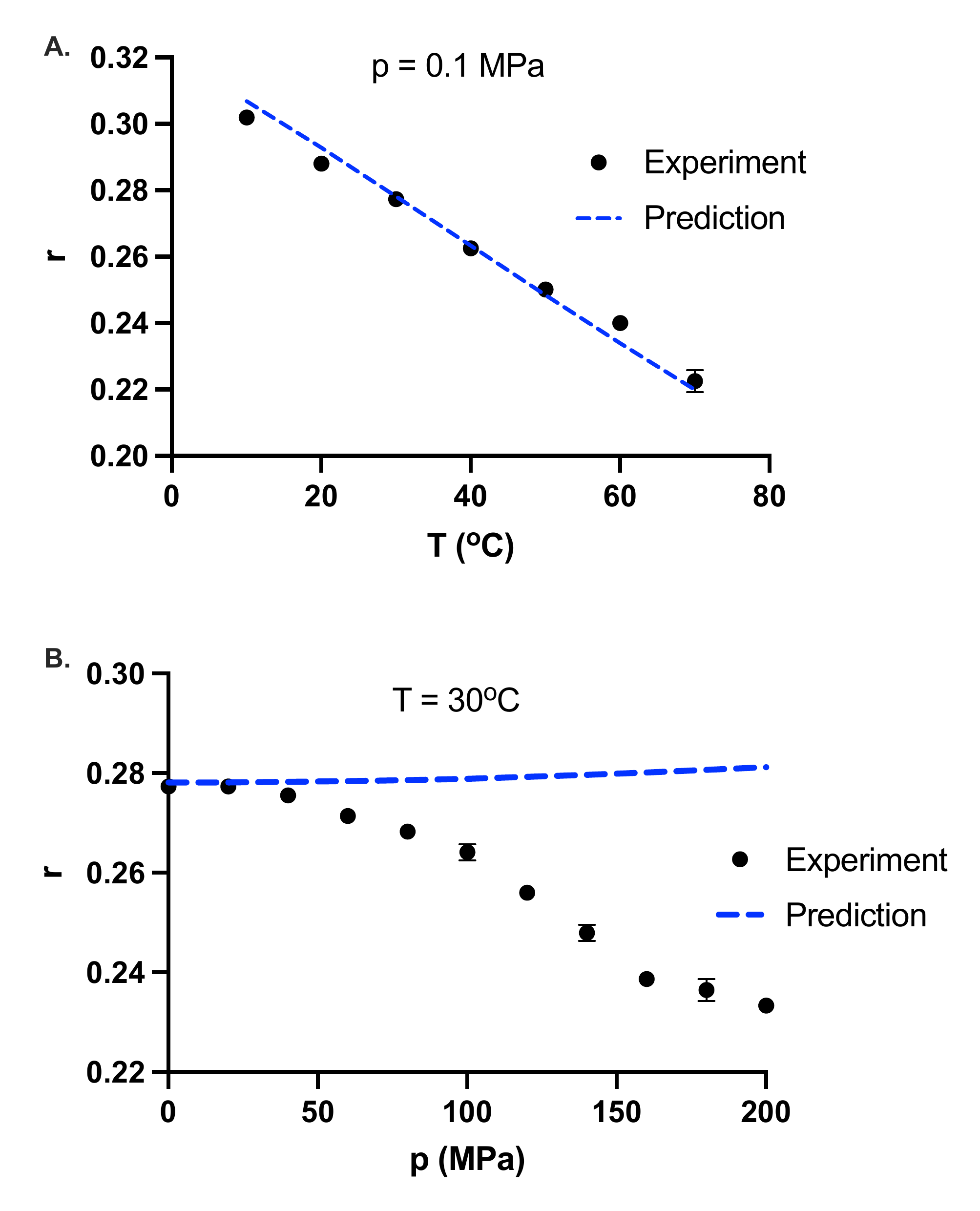}
 \end{center}
 \caption{Comparison between experimentally measured and predicted values of the fluorescence anisotropy taking into account of pressure and temperature dependence of the dynamic viscosity of the solvent and calculated using Eq.~\ref{eq:eqDebyePerrin}. (A) Temperature dependence of experimentally observed anisotropy (solid black circles) and predicted values of  anisotropy (dotted blue curve) at $P=0.1$~MPa. Anisotropy decreases with increasing temperature. (B) Pressure dependence of experimentally observed anisotropy (solid black circles) and predicted values of  anisotropy (dotted blue curve) at $T=30^o$~C. In contrast to experiments, the predicted values of fluorescence anisotropy of GFP increase slightly with increasing pressure.}
\label{fig:fig6}
\end{figure}

We next investigated the extent to which the pressure-temperature induced variations in viscosity of the solvent affect the changes observed in anisotropy. The rotational correlation time, $\tau_R(p,T)$, is related to the pressure-temperature dependent viscosity, $\eta(p,T)$, of the solvent and the temperature, and is given by the Debye-Stokes-Einstein relation
\begin{equation}
\tau_R(p,T) = \frac{V\eta(p,T)}{k_BT},
\label{eq:eqDebye}
\end{equation}
where $V$ is the hydrodynamic volume of the molecule, and $k_B$ is the Boltzmann constant. The Perrin equation (Eq.~\ref{eq:eqAniso}) can now be written as 
\begin{equation}
r(p,T)= \frac{r_0}{1+\frac{\tau_Fk_BT}{V\eta(p,T)}}
\label{eq:eqDebyePerrin}
\end{equation}
At room temperature and atmospheric pressure, the fluorescence lifetime and the effective hydrodynamic radius of GFP are $2.5$~ns and $2.21$~nm, respectively~\cite{hink2000structural}. Since a change in viscosity by itself can give rise to a change in rotational correlation time, and therefore, will result in a change in anisotropy. The viscosity of the solvent in the limit of dilute concentration of salts is approximately the viscosity of water. We computed the values of dynamic viscosity of water using International Association for the Properties of Water and Steam (IAPWS)  formulation~\cite{cooper2008release} for all the pressures and temperatures studied here. 
It is well established that hydrophobicity of amino acids decreases at high pressure. Indeed, the high-pressure denaturation of proteins precedes with penetration of water molecules inside the protein~\cite{hummer1998pressure}, leading to swelling and further exposure of hydrophobic residues to water~\cite{wolfenden1994probability}. This suggests that the hydrodynamic volume of the protein may increase with increasing pressure resulting in larger correlation time and hence larger anisotropy. However, the pressure dependence of the hydrodynamic radius of  proteins is not always monotonic~\cite{bohidar1989light, chryssomallis1981effect}. For example, in case of lysozyme, hydrodynamic radius of the protein does not change up to $p=100$~MPa, decreases by about $3$\%  between $120$~MPa to $230$~MPa, and increases by about $20\%$ above $600$~MPa~\cite{chryssomallis1981effect}. Even if it was the case that hydrodynamic volume decreases with pressure, one can not expect a 83\% change in the volume of GFP  to compensate for the anisotropy decrease with pressure at $T=30^{\circ}$C. It is safe to assume that the size of the GFP does not change significantly up to $P=200$~MPa. To test whether the decrease in anisotropy with pressure can be ascribed to rotational correlation time changes of protein, we next calculated the values of $\tau_R$ for different state points using Eq.~\ref{eq:eqDebye} and assuming that the hydrodynamics volume of GFP does not change in the range of pressure ($0.1-200$~MPa) and temperature ($10-70^{\circ}$C). Using Eq.~\ref{eq:eqDebyePerrin} and assuming a constant $\tau_F$, one can estimate the values of anisotropy due to pressure-temperature induced changes in viscosity. In Fig.~\ref{fig:fig6}(A), we compare the temperature dependence of predicted anisotropy using Eq.~\ref{eq:eqDebyePerrin} with experimental values of anisotropy at $p=0.1$~MPa. Since the solvent viscosity decreases with temperature, the predicted anisotropy also decreases with temperature. In Fig.~\ref{fig:fig6}(B), we compare the pressure dependence of predicted anisotropy and experimental values of anisotropy at a constant temperature $T=30^{\circ}$C. Weak dependence of predicted anisotropy with increasing pressure is due to a weak dependence of viscosity of solvent with pressure. In contrast to temperature dependence (Fig.~\ref{fig:fig6}(A)), where the predicted and experimental values are similar, the pressure dependence of experimental values of anisotropy deviates a lot from the predicted anisotropy. While the pressure dependence of the viscosity predicts a slight increase in anisotropy with increasing pressure, the experimental values of anisotropy decrease sharply with pressure. This suggests that the observed decrease in anisotropy upon increasing pressure cannot be attributed to viscosity changes. 
 \begin{figure*}
 \begin{center}
\includegraphics[width=14cm]{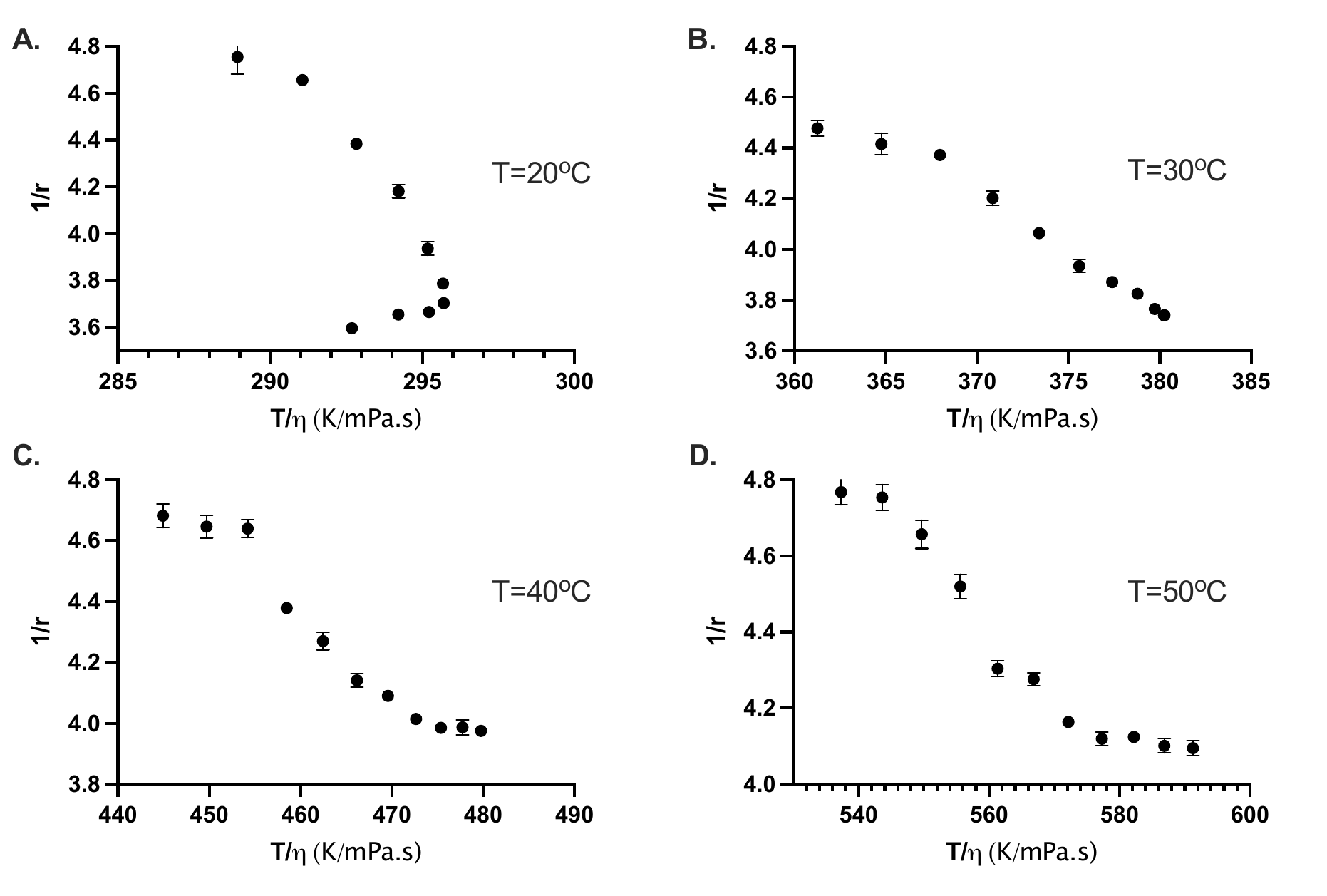}
\end{center}
\caption{Perrin-Weber plot of $1/r$ as a function of $T/\eta$ for (A) 20$^{\circ}$C , (B) 30$^{\circ}$C , (C) 40$^{\circ}$C , and (D) 50$^{\circ}$C. The error on $1/r$ were estimated from the error on $r$. Error bars where not visible are smaller than the symbol size.}
\label{fig:figperrin}
\end{figure*}
However, the GFP chromophore may enjoy its own local mobility that may not be coupled to the protein's rotation~\cite{jameson2010fluorescence}. In that case, it is informative to check Perrin-Weber plot of $1/r$ as a function of $T/\eta$. When the chromophore's mobility is strongly coupled to the viscosity, $1/r$ is positively correlated and varies linearly with $T/\eta$. When the probe enjoys local mobility, $1/r$ can become a non-linear function of $T/\eta$~\cite{wahl1967fluorescence} and an increasing slope at large values of $\eta$ (smaller values of $T/\eta$) would mean larger local mobility~\cite{wahl1967fluorescence,jameson2010fluorescence,vandermeulen1990excitation}. In Fig.~\ref{fig:figperrin}, we show $1/r$ as a function of $T/\eta$ for four different temperatures. The viscosity $\eta(p,T)$ were calculated as described above. We find that $1/r$ is non-linear function of $T/\eta$ for all the temperatures studied here, Moreover, for $T=20^{\circ}$C, $1/r$ displays a positive correlation for $p\le80$~MPa and negative correlation for larger pressures, resulting in a nose-shaped curve due to the viscosity anomaly of water~\cite{bett1965effect,prielmeier1987diffusion,schmelzer2005pressure}. For $T\ge 30^{\circ}$C,  $1/r$ decreases monotonically with $T/\eta$. It is interesting to note that for low temperature ($T<30^{\circ}$C), $1/r$ is positively correlated with $T/\eta$ up to the pressure where the viscosity anomaly of the water disappears. Viscosity of diffusion anomaly in water at low temperatures has been attributed to breaking of hydrogen bonds with pressure and subsequent saturation of disordered hydrogen bond network at higher pressures~\cite{singh2017pressure,okada2005pressure}. It is likely that H-bonding environment around the chromophore is affected at high pressure resulting in the changes in radiative and non-radiative decay channels.
 \begin{figure*}
\begin{center}
 \includegraphics[width=14cm]{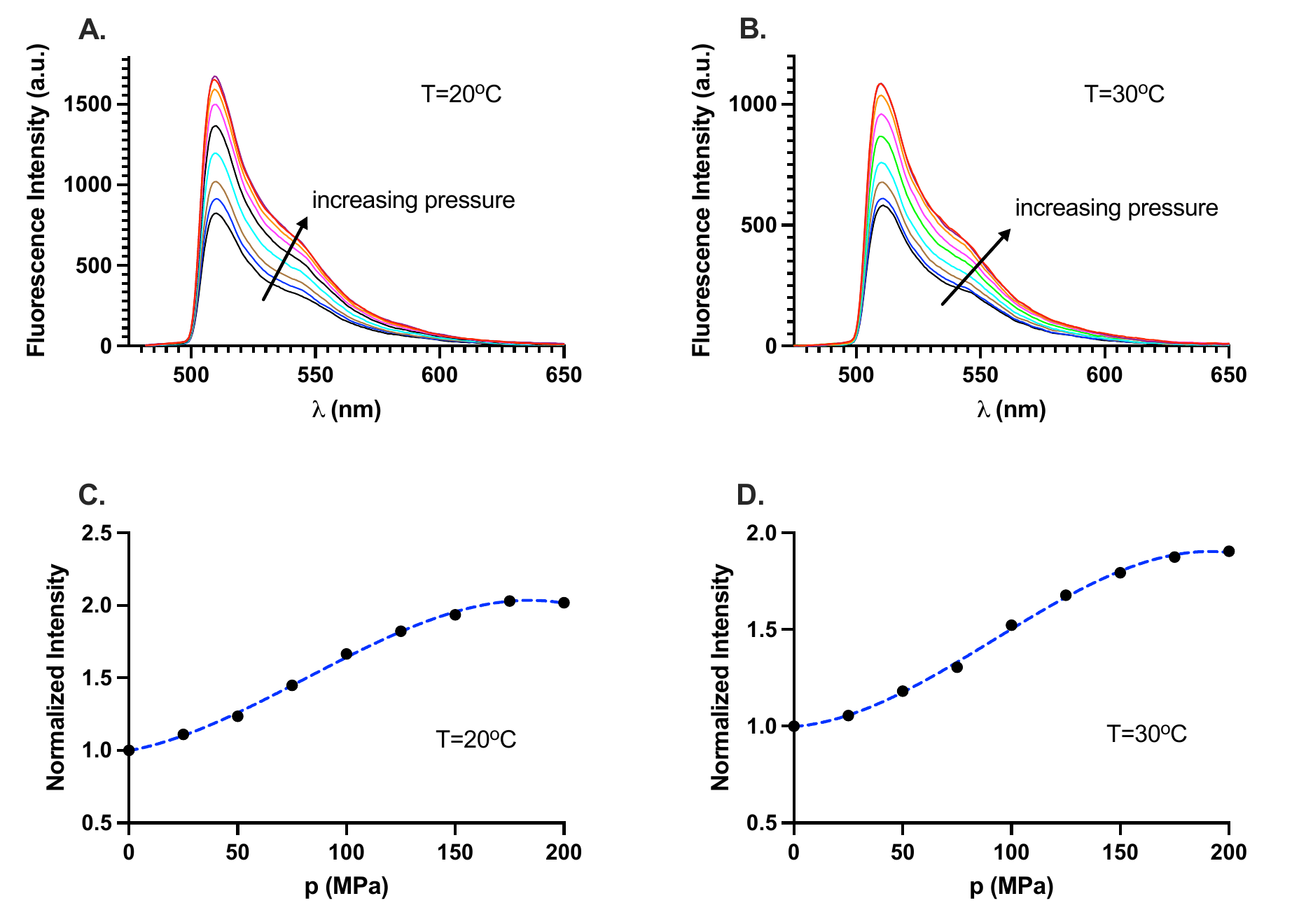}
\end{center}
\caption{Fluorescence intensity as a function of wavelength for different pressures at (A) $20^{\circ}$C, and (B) $30^{\circ}$C. Normalized fluorescence intensity as a function of pressure at (C) $20^{\circ}$C, and (D) $30^{\circ}$C. Dotted blue curves are the polynomial fits through the data points and only serve as the guides to the eye.}
\label{fig:fig9}
 \end{figure*} 
Decrease of anisotropy with pressure can arise due to increase in fluorescence lifetime resulting from changes in the radiative and non-radiative decay channels. Quantum yield, $\Phi$, depends on the both the radiative and non-radiative decay rates as
\begin{equation}
\Phi = \frac{k_r}{k_r + k_{nr}}
\end{equation}
where $k_r$ and $k_{nr}$ are radiative and non-radiative decay rates, respectively. High pressure may affect both radiative and non-radiative decay rates, and, therefore the quantum yield. If the ratio $\frac{k_{nr}}{k_r}$ decreases, quantum yield increases. To test whether quantum yield changes upon increase of pressure, we obtained fluorescence spectra of GFP as a function of pressure for two different temperatures, T=$20^{\circ}$C and $30^{\circ}$C at $488$~nm excitation. In Figs.~\ref{fig:fig9} (A) and~\ref{fig:fig9}(B), we show the fluorescence spectra of GFP for different pressures at $T=20^{\circ}$C and $T=30^{\circ}$C, respectively. We find that the fluorescence intensity increases monotonically with pressure in the entire emission wavelength range. In Figs.~\ref{fig:fig9}(C) and~\ref{fig:fig9}(D), we show the integrated fluorescence intensity relative to the integrated fluorescence intensity at $P=0.1$~MPa as a function of pressure. We find the total fluorescence intensity increases by a factor of $~2$ at highest pressure studied here. We note that we do not measure the quantum yield directly but an increase in the fluorescence could be suggestive of increase in quantum yield with pressure. Furthermore, our experiments do not resolve the questions of how individual radiative and non-radiative decay processes are affected at high pressure.

A decrease in anisotropy at high pressures can also result from other sources including an increase of the fluorescence lifetime of GFP due to change in refractive index. The fluorescence lifetime of GFP decreases as the inverse of the square of refractive index~\cite{suhling2002influence,tregidgo2008effect}. The refractive index of the TMD buffer used in our experiments is similar to the refractive index of pure water due to low salt concentration ($10$~mM). Indeed, the refractive index of dilute aqueous solution of salts is not much different from pure water~\cite{aly1993refractive,leyendekkers1977refractive}. Refractive index of pure water increases with increasing pressure~\cite{weiss2012water}, and therefore, the fluorescence lifetime should decrease with pressure~\cite{suhling2002influence,tregidgo2008effect}. However, decrease of fluorescence lifetime cannot lead to a decrease of fluorescence anisotropy with pressure.

\section{Discussion}
We have measured fluorescence anisotropy  of GFP in a wide range of pressure ($0.1-200$~MPa) and temperature ($10-70^{o}$C). At room temperature and atmospheric pressure, we find that the anisotropy of GFP in Tris-HCl buffer is about $0.28$,  which is similar to the value of anisotropy of GFP ($\approx0.27$) in Phosphate Buffer Saline reported elsewhere~\cite{donner2012mapping}. This high value of the anisotropy of GFP is due to short fluorescence lifetime($\approx2.5$~ns) as compared to its rotational correlation time ($\approx10$~ns)~\cite{hink2000structural}.  
Our results show that  the fluorescence anisotropy of the GFP decreases with temperature at atmospheric pressure but exhibits a maximum with temperature for $P\geq20$~MPa. Furthermore, we find that at a constant temperature, anisotropy decreases sharply with increasing pressure. The isoanisotropic contours in pressure-temperature plane exhibit elliptic shapes, typical of pressure-temperature stability phase diagrams of protein. This suggests that the anisotropy of GFP reflects the stability of protein with pressure and temperature. 

A decrease in anisotropy with temperature at constant pressure can be partly attributed to the temperature dependence of the viscosity of the solvent. Experimental observation of the decrease in anisotropy upon increasing pressure cannot be ascribed to viscosity changes. A decrease in anisotropy at high pressures can result from other sources including a decrease in rotational correlation time and an increase of the fluorescence lifetime of GFP. We find that the thermal stability of GFP increases with pressure. Moreover, Perrin-Weber plots suggest that 

There is a large body of work that suggest that local environment around the chromophore may affect its fluorescence lifetime. Variation in pH~\cite{heikal2001multiphoton}, viscosity~\cite{suhling2002influence}, temperature~\cite{mauring2005enhancement} and pressure~\cite{foguel1992pressure} have been shown to have an effect on the fluorescence lifetime of different fluorophores. Although, viscosity of the solvent does seem to affect the fluorescence lifetime, there is no correlation between the viscosity and the fluorescence lifetime of GFP~\cite{suhling2002influence}. Dependence of fluorescence lifetime  on refractive index of the solvent has been investigated by Suhling et. al.~\cite{suhling2002imaging}. These authors find that the inverse of the fluorescence lifetime scales linearly with square of the refractive index for GFP~\cite{suhling2002imaging,tregidgo2008effect,davidson2020measurement} and Enhanced Cyan and Yellow Fluorescent Protein~\cite{borst2005effects}. At a given temperature, the refractive index of water increases with increasing pressure~\cite{waxler1963effect}. However, these changes in refractive index would only lead to decrease in fluorescence lifetime.

 To this end, we also note that Mauring et. al~\cite{mauring2005enhancement} observed fluorescence enhancement of the blue fluorescent protein (BFP) with hydrostatic pressure. The coupling of BFP chromophore with the rest of the protein is different as compared to GFP because of the His66 substitution, which leads to smaller number of hydrogen bonds~\cite{heim1996engineering, heim1994wavelength,megley2008photophysics}. Due to smaller number of hydrogen bonds, the fluorescence lifetime and the quantum yield of BFP is much smaller compared to GFP. They find that the fluorescence quantum yield increases with pressure without a change in the shape of emission spectra, results very similar to our results for GFP. They further attribute the increase in fluorescence quantum yield with pressure to the inhibition of fast quenching processes due to stabilization of hydrogen bond between the chromophore and the rest of the protein. We expect that the decrease in anisotropy may reflect the effect of pressure on the radiative and non-radiative decay processes arising due to pressure-temperature effect on the H-bonding environment around the chromophore. Our current experimental setup does not allow us to perform time-resolved experiments, which on the other hand, would have been an ideal way to resolve these issues, and should be explored in future.

\section*{Author Contributions}
PK conceived and designed the research, HK, KN, and PK performed the research, analyzed the data, and wrote the paper. 

\section*{Conflicts of interest}
There are no conflicts to declare.

%%%REFERENCES%%%
\bibliography{knk} %You need to replace "rsc" on this line with the name of your .bib file
\bibliographystyle{rsc} %the RSC's .bst file

\end{document}